\documentclass[12pt]{iopart}

\expandafter\let\csname equation*\endcsname\relax
\expandafter\let\csname endequation*\endcsname\relax
\usepackage{amsmath}
\usepackage{graphicx,color,cite}
\usepackage[colorlinks=true,linkcolor=red,urlcolor=blue,citecolor=blue]{hyperref}

\begin{document}

\title{Interspecies entanglement with impurity atoms in a lattice gas}

\author{S~Sarkar$^{1,2,3}$, S~McEndoo$^3$, D Schneble$^4$ and A~J~Daley$^3$} 

\address{Institute of Physics, Polish Academy of Sciences, Aleja Lotnikow 32/46, PL-02668 Warsaw, Poland}
\address{Harish-Chandra Research Institute, HBNI, Chhatnag Road, Jhunsi, Allahabad 211 019, India}
\address{Department of Physics and SUPA, University of Strathclyde, Glasgow G4 0NG, UK}
\address{Department of Physics and Astronomy, Stony Brook University, Stony Brook, New York 11794-3800, USA}

\ead{sarkar@ifpan.edu.pl}

\begin{abstract}
The dynamics of impurity atoms introduced into bosonic gases in an optical lattice have generated a lot of recent interest, both in theory and experiment. We investigate to what extent measurements on either the impurity species or the majority species in these systems are affected by their interspecies entanglement. This arises naturally in the dynamics and plays an important role when we measure only one species. We explore the corresponding effects in strongly interacting regimes, using a combination of few-particle analytical calculations and Density Matrix Renormalisation group methods in one dimension. We identify  how the resulting effects on impurities can be used to probe the many-body states of the majority species, and separately ask how to enter regimes where this entanglement is small, so that the impurities can be used as probes that do not significantly affect the majority species. The results are accessible in current experiments, and provide important considerations for the measurement of complex systems with using few probe atoms.

\end{abstract}


\maketitle

\section{Introduction}
\label{Introduction}

Over the past two decades, there has been much progress in understanding the rich physics of impurity atoms introduced in to Bose gases \cite{Bardeen1967,Astrakharchik2004,Klein2005}. The resulting phenomena ranges from the realisation of models of open quantum systems \cite{Recati2005,Haikka2011,Haikka2012,Haikka2013,Mcendoo2013} and mediated interactions \cite{Klein2005} to impurity dynamics \cite{Palzer2009,Bonart2012} and broader studies of polarons \cite{Bei-Bing2009,Tempere2009,Casteels2013,Stojanovic2012,Blinova2013}, where introduction of the  impurity gives rise to a collective object incorporating localised wavepackets of excitations in the Bose gas, with an increased effective mass \cite{Feynman1954,Miller1962,Astrakharchik2004}. There has been a lot of recent theoretical work on these systems, applying new variational and field theory techniques to the problem across a variety of parameter regimes \cite{Rath2013,Benjamin2014,Christensen2015,Grusdt2015,Kain2016,Levinsen2015,Parisi2017,Shchadilova2016,Shchadilova2016b,Vlietinck2015,Volosniev2017,Li2014,Grusdt2017,Dehkharghani2015,Ardila2015,Ardila2016,Camacho-Guardian2018,Charalambous2018}. There has also been extensive experimental progress in observing polarons in Bose gases \cite{Bloch2008a,Palzer2009,Catani2012,Gadway2010,Jorgensen2016,Scelle2013,Rentrop2016}, also in a strongly interacting regime \cite{Hu2016,Yan2020}.

At the same time, entanglement in many-body systems \cite{Amico2008,Eisert2010} has generated a lot of interest, especially because of the information that can be extracted from entanglement in spatial modes. This is helpful in understanding a variety of phenomena, e.g., to identify  topological phases \cite{Kitaev2006,Levin2006,Jiang2012,Isakov2011} or understand the growth of local entropy during thermalisation \cite{Kaufman2016,Rigol2008}. Such entanglement in space has also been measured in experiments with cold atoms in optical lattices \cite{Islam2015,Daley2012,Moura-Alves2004}. For multicomponent gases, the interplay between correlation and entanglement effects has been explored in both the Bose-Hubbard and the Hubbard model \cite{Coe2010,Keiler2018,Lingua2016, Richaud2020}. Recently, interspecies entanglement of has been used to characterise the shift of the phase transition points in the two-component Bose Hubbard model due to interspecies interactions \cite{wei2016}. Entanglement of an impurity in a few-body continuous system has also been discussed in terms of a reduced single-particle density matrix in Ref.~\cite{Garcia-March2016}.

The system we will investigate here is a lattice model for impurities introduced into a Bose gas loaded into the lowest Bloch band of an optical lattice \cite{Bruderer2007,Bruderer2008,Massel2013}. This has been realised in experiments \cite{Gadway2010,Fukuhara2013a,Chen2014}, and recently discussed as an important example for characterising the probing  of strongly correlated systems with impurity atoms~\cite{Cosco2018}. We are particularly interested in asking about the role of entanglement between the impurity atoms and the majority species, and how this impacts measurements made on a single species alone. Specifically, when momentum distributions of impurity atoms are measured in some two-species experiments \cite{Gadway2010,Catani2012} there is a notable decrease in the visibility of peaks in these momentum distributions, beyond what might be expected from an increase in the effective mass or interactions mediated by the majority species.  Entanglement between two species can lead to a mixed reduced density operator for a single species, substantially affecting first-order coherence properties, including measured momentum (or quasi-momentum) distributions. These effects, as we will see below, are strongest when going away from the limit where polaron physics is studied (or indeed, where a usual polaron description is valid), i.e., it is important for strong interactions, or when there is not a large ratio between the atom numbers of the majority and minority species.

\begin{figure}[t]
\hspace{1in} 
\includegraphics[width=0.75\linewidth]{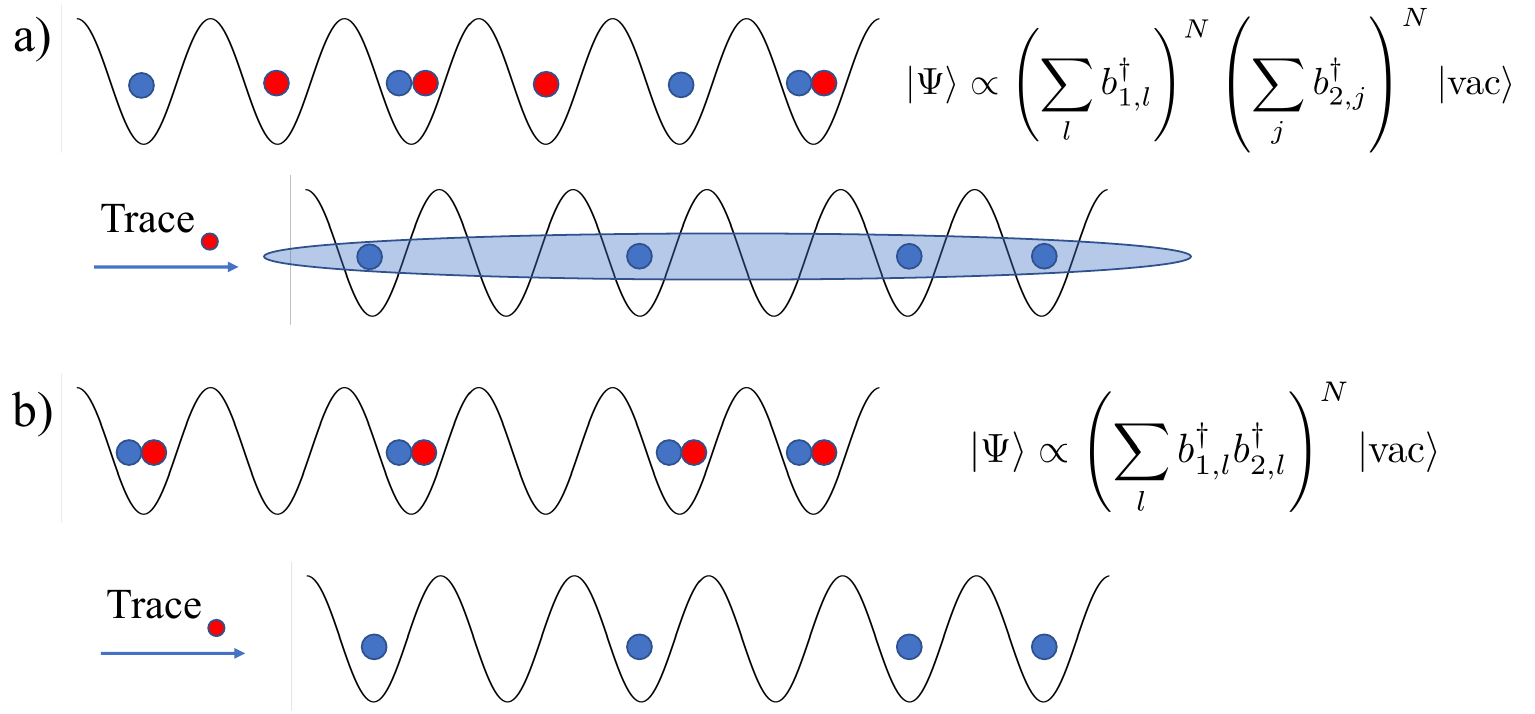}
\label{figure1}\caption{Schematic depiction of the effects of entanglement between two interacting species (red and blue) on a lattice, when we effectively trace over the red species in a measurement of the blue species. (a) In the case with no interspecies interaction (here $N=4$ non-interacting particles of each species), when we trace out the red species, the state of the remaining species remains a pure state, corresponding to a coherent superposition of all configurations of the particles delocalised on the lattice. (b) In contrast, starting from a state of delocalised red-blue dimers, when we trace out the red species, the blue species is left in a mixed state of all possible arrangements of the particles on the lattice.}
\end{figure}

As an example, in figure~\ref{figure1} we depict a two-component mixture in a lattice. In figure~\ref{figure1}(a), the two species are in a product state with no entanglement, as might be expected to occur in the ground state of a mixture with no interspecies interaction. If we trace over one species and ask what the reduced density operator is that describes the second species alone, we find it represents a pure state, with the atoms still delocalised, and with their momentum peaked at $p=0$. However, in figure~\ref{figure1}(b), we have an initial state of perfectly correlated dimers between the two species, with dimer momentum peaked at $p=0$. This represents an entangled state of the two species, and when we now trace over one species, the reduced density operator for the other species contains a mixture of all possible configurations, with a completely flat momentum distribution. 

Below, using numerical and analytical methods, we analyse this behaviour for different parameter regimes of both impurity and majority species particles. We find complex behaviour that exhibits particular signatures associated with the quantum phase diagram of the Bose-Hubbard model for the majority species. In this sense, understanding the impurity-majority species entanglement can be useful as an alternative route to probing the complex many-body behaviour, either by observing the impurity atoms, or by observing their effect on the state of the majority atoms. At the same time, if the impurity atoms are being used as a probe in the sense of Refs.~\cite{Mcendoo2013,Cosco2018}, it may be important to minimise direct entanglement between the species, and we analyse requirements in simple example cases to achieve this regime.  

The remainder of this article is organised as follows: In section~\ref{model} we introduce the model and numerical methods we use to analyse the system. We then analyse the entanglement and momentum distributions for two particles on a lattice, one of each species, as a starting point for further investigations in section~\ref{Two particles on 1D lattice}. In section~\ref{Comparison}, we make use of a Born-Oppenheimer approximation to help us separate, for simple examples, the effects of a change in effective mass of the impurities from effects arising from the entanglement between the impurities and the majority species. In section~\ref{Many body problem} we analyse the behaviour for systems of multiple atoms on the lattice in parameter regimes corresponding to different quantum phases of the majority species. In section~\ref{BEC} we then ask under which circumstances a single atom will become disentangled from the system to which it is coupled, and to this end we investigate cases where the impurity atoms and the majority species have different tunnelling rates in the lattice, and where the impurity atoms are confined to a fraction of the full length of the system. We then provide a conclusion and outlook in section~\ref{Conclusion}. 

\section{Model}
\label{model}

We consider an ensemble of bosonic atoms loaded into the lowest band of an optical lattice. We denote the majority species as species $1$, with $N_1$ atoms, and the impurity species as $2$, with $N_2\leq N_1$ atoms. For sufficiently low temperatures and where interactions are smaller than the energy separation between Bloch bands, this situation is generally well described by a multi-species Bose-Hubbard model ($\hbar\equiv 1$) \cite{Bloch2008a},
\begin{align} 
H_{BH}= -\sum_{\langle l,j \rangle , \sigma} J_{\sigma}b_{\sigma,l}^{\dag}b_{\sigma,j}+\displaystyle\sum_{l,\sigma }\frac{U_{\sigma}}{2}n_{\sigma,l}(n_{\sigma,l}-1) +\displaystyle\sum_{l}U_{12}n_{1,l}n_{2,l}
\,, \label{HBH} \end{align}
where $b_{\sigma,i}^{\dag}\left(b_{\sigma,l}\right)$ and $n_{\sigma,l}$ are the creation(annihilation) operator and number operator for species $\sigma\in \left\{1,2\right\}$ on the $l$-th lattice site. Each species has nearest-neighbour tunnelling rate $J_\sigma$ and intra-species onsite interaction $U_\sigma$. The on-site interspecies interaction energy shift is then denoted $U_{12}$. We will generally take $J_1=J_2\equiv J$ unless otherwise specified, and we usually take the same 1D lattice length $M$. In section~\ref{BEC} we will consider the impurity particles to be confined to a lattice length $M_2\leq M$, and denote $M_1$ as the full length for the majority species.

Throughout this work we will mainly restrict our calculations to one dimension, in order to simplify the computations. However the basic principles we discuss here and the qualitative behaviour of the entanglement in different parameter regimes is expected to transfer directly to higher dimensions. 

In what follows, we will use analytical methods to obtain exact results for a few atoms, and exact diagonalisation methods for small lattice sizes, especially to obtain values for the von Neumann entropy of entanglement. If we compute the reduced density matrix $\rho_\sigma$ for either of the two species, 
\begin{align}
\rho_\sigma={\rm Tr}_{\overline{\sigma}}\{|\Psi\rangle\langle \Psi | \},
\end{align}
where $|\Psi\rangle$ denotes the state of the total system, and $\overline{\sigma}$ here denotes the opposite species to $\sigma$, then the von Neumann entropy of entanglement can be computed as
\begin{align}
S_{\text{vN}}=-\text{Tr}\{\rho_\sigma\text{log}_2\rho_\sigma\}.
\end{align}
Note that if the total state of the system is pure, then $S_{\text{vN}}$ is independent of the choice of $\sigma$. The entropy of the reduced density matrix for one species in this case entirely represents the entanglement between the two species. If $S_{\text{vN}}=0$ the reduced density matrix is a pure state. This occurs when the entanglement is zero, and the total state is a product state of the two species.

As a guide to larger system behaviour, we employ mean-field methods based on the bosonic Gutzwiller ansatz \cite{Rokhsar1991}, where the ground state of the two species Bose-Hubbard Hamiltonian in \eqref{HBH} on an $M$-site chain is written as, 
\begin{align}
|\psi\rangle=\prod_{l=1}^M \sum_{n_1,n_2}\frac{f_{n_1,n_2}^{(l)}}{\sqrt{n_1!n_2!}}(b_{1,l}^\dagger)^{n_1}(b_{2,l}^\dagger)^{n_2}|{\rm vac}\rangle
\,.\end{align}   
Here $f_{n_1,n_2}^{(l)}$ is the amplitude associated with different number states for each particle on the $l$-th site. 

To provide additional information on the many-body physics beyond this, we employ density matrix renormalisation group (DMRG) methods based around matrix product states \cite{White1992,Schollwock2011} to determine the ground state. In each case, we ensure that the results are properly converged in the bond dimension of the matrix product state, $D$. 

\section{Two particles on a 1D lattice}
\label{Two particles on 1D lattice}

We can obtain an intuition for the behaviour we expect by considering the entanglement of two distinguishable particles in an optical lattice. Using an exact solution for the ground state of Eq.~\eqref{HBH} \cite{Wouters2006,Winkler2006,Valiente2008}, we can quantify how the entanglement and momentum distribution for each particle change as a function of the interaction strength between the bosons. 

As a useful starting point, we consider the limiting cases. Because the ground state of a single particle on the lattice is a state with quasi-momentum $p=0$, for two particles the non-interacting ground state when $U_{12}=0$ is given by
\begin{align}
|\Psi_{\rm prod}\rangle = \frac{1}{M} \left(\sum_l b_{1,l}^\dagger\right) \left(\sum_{l'} b_{2,l'}^\dagger\right)  |{\rm vac}\rangle. 
\end{align}
When we compute the reduced density matrix, we then obtain
\begin{align}
\rho_1={\rm Tr}_2\left \{|\Psi_{\rm prod}\rangle\langle \Psi_{\rm prod}|\right\}= \frac{1}{M} \left(\sum_l b_{1,l}^\dagger\right) |{\rm vac}\rangle\langle {\rm vac}| \left(\sum_{l'} b_{1,l'}\right),
\end{align}
which is a pure state, for which the resulting quasi-momentum distribution $n(p)$ is peaked at $p=0$. If, however, the particles are interacting such that $|U_{12}| \gg J$, then for attractive interaction, we obtain instead 
\begin{align}
|\Psi_{\rm ent}\rangle = \frac{1}{\sqrt{M}}\sum_l b_{1,l}^\dagger b_{2,l}^\dagger |{\rm vac}\rangle. 
\end{align}
If we now compute the reduced density matrix, we then obtain
\begin{align}
\rho_1=\frac{1}{M}\sum_l \left( b_{1,l}^\dagger |{\rm vac}\rangle\langle {\rm vac}| b_{1,l}\right),
\end{align}
which is a mixed state with $S_{\text{vN}}=\log_2 M$. This mixed state with particles localised on each site arises in a sense because the state of the second species contains information on the locations of the state of the first species. The resulting momentum distribution is completely flat, despite the fact that the doublon momentum distribution is peaked at $p=0$.

To analyse the behaviour for arbitrary interaction strengths we look at the general solution for the two-particle wavefunction
\begin{align}
|\Psi_2\rangle=\sum_{x,y}\psi(x,y)b_{1,x}^\dagger b_{2,y}^\dagger |{\rm vac}\rangle,
\end{align}
with complex coefficients $\psi(x,y)$. Taking periodic boundary conditions, we can separate the centre of mass $R=(x+y)/2$ and relative $r=x-y$ coordinates, and determine an analytical solution \cite{Wouters2006,Winkler2006,Valiente2008}, for which we provide more details in \ref{appendix_two}.

\begin{figure}[t]
\hspace{1in}
\includegraphics[width=0.75\linewidth]{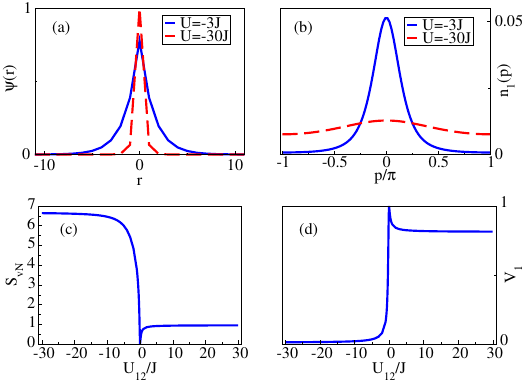}
\label{figure2} \caption{\label{two} Entanglement and momentum distribution visibility for two distinguishable particles on a 1D lattice of $M=101$ sites with periodic boundary conditions. (a)  The pair wavefunction $\psi(r)$ for attractive interaction where $r$ is the inter-particle separation. Here we plot the two cases where the interaction $U=-3J$ and $U=-30J$. (b) The corresponding momentum distributions for $U=-3J$ and $U=-30J$. (c) The single particle von Neumann entropy $S_{\text{vN}}$, showing the entanglement between the two particles as a function of interaction strength $U_{12}$. (d) The single particle visibility $V_1$, defined as the height of the momentum distribution peak, as a function of interaction $U_{12}$.}
\end{figure}

We show the analytical calculation for this relative wavefunction in figure~\ref{two}(a). We see that the peak of the bound pair solutions becomes sharper as the interaction strength is increased. The single particle momentum distributions are shown for comparison in figure~\ref{two}(b), and show clearly the effect of interactions. As expected for $U<0$, with increasing interaction strength the two particles become more tightly bound and this leads to broadening of the single-particle momentum distribution. In this general setting, the entanglement between the two particles causes mixedness in the single species reduced density matrix, implying a reduction in the first-order coherence of this species when measured alone. Therefore, the momentum distribution, although being a single particle observable, is affected by the entanglement between species. In the rest of the article we will use the height of the momentum peak or \textit{visibility} for each species $\sigma$, $V_\sigma$, as an indicator of the corresponding changes in momentum distributions. 

In figure~\ref{two}(c) we then look directly at the von Neumann entropy of entanglement. For attractive interactions ($U_{12}<0$), the entanglement grows very sharply as a consequence of the direct pairing of the particles in position space that creates the bound state and reaches the saturation value for small $U_{12}/J$ (which is $\log_2 M$ as noted above). The entanglement in position space is generated by repulsion ($U_{12}>0$) between the particles, which makes it energetically unfavourable for them to be present on the same lattice site. This does not provide as strong entanglement as in the attractive case, but still increased with increasing $U_{12}$, towards an asymptotic value. The visibility of the momentum distribution peak $V_1$ is shown in figure~\ref{two}(d) and directly mirrors the behaviour of the entanglement, falling very sharply on the attractive side $U_{12}<0$ as the particles become highly entangled and the momentum distribution for a single particle tends rapidly to a flat distribution. Similarly, on the repulsive side ($U_{12}>0$), the slight drop of the visibility profile followed by a steady value reflects the corresponding entanglement of the particles. 

\section{Comparison of the effects of entanglement with mediated interactions and increased effective mass}

\label{Comparison}

\begin{figure}[t]
\hspace{1in}
\includegraphics[width=0.75\linewidth]{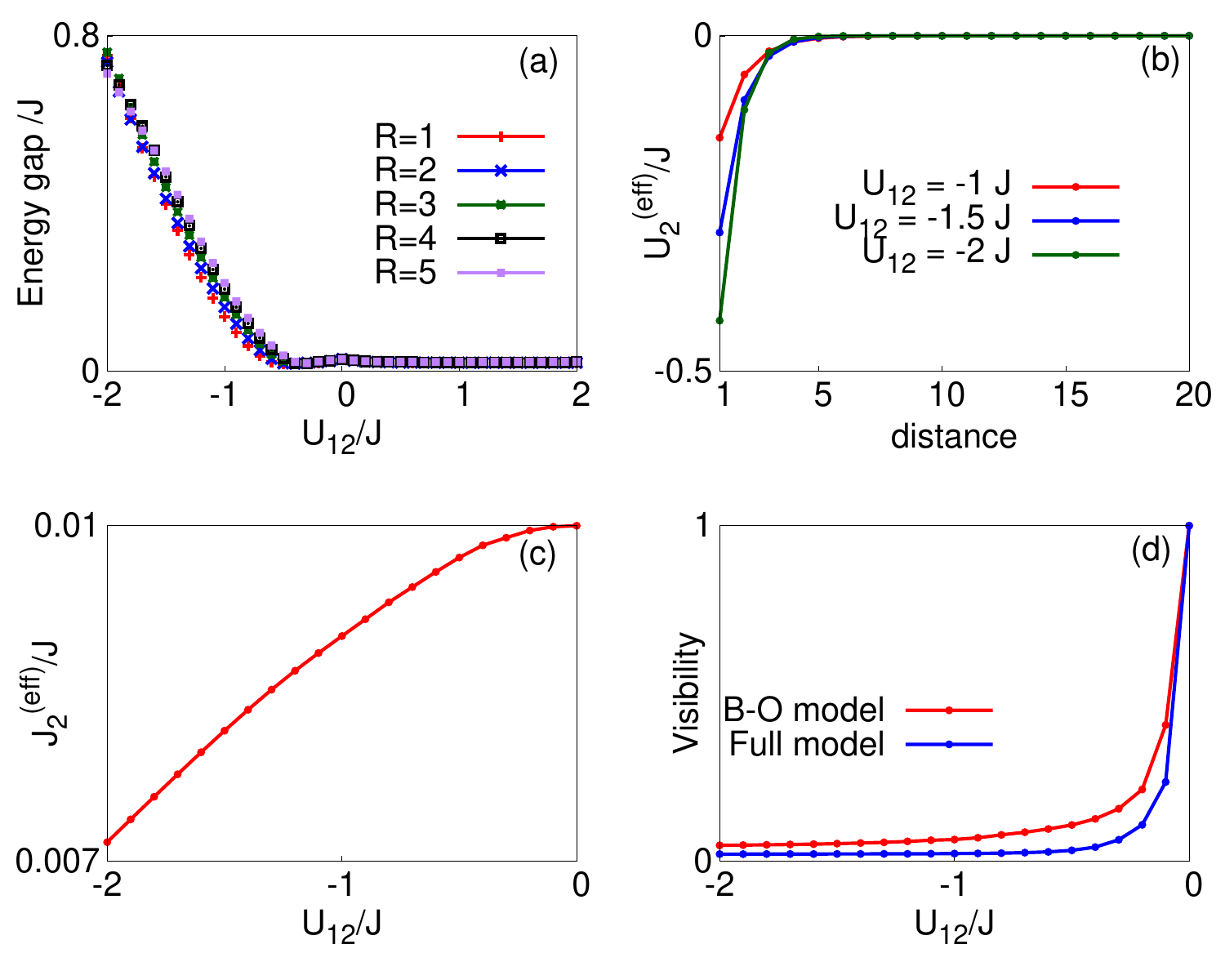}
\caption{(a) The energy gap between the ground state and the first excited state, for various distances $R$ between the two impurities, for $50$-site chain with two bosons from each species, and $U_1=2J$. The B-O approximation is valid for sufficiently attractive $U_{12}$ and is expected to break down for repulsive $U_{12}$ and close to $U_{12}=0$. (b) The induced long-range interaction $U_2^{\text{eff}}$ between the two impurities calculated for various $U_{12}$ values with $U_1=2J$. (c) The effective tunnelling rate $J_2^{\text{eff}}$ for an impurity boson with $J_2=0.01J$. (d) Visibility of peaks in the momentum distribution for the impurity species with interspecies interaction $U_{12}$, shown separately from the model with extracted effective interactions and tunnelling rates in the Born-Oppenheimer approximation (red), and for the full model (blue). \label{fig_BO_1}}
\end{figure}

In some sense, this decrease in the visibility of a momentum distribution for impurity atoms, when interacting with a majority species, might be expected. After all, the majority species will tend to mediate additional interactions between the minority atoms, and also potentially to increase the effective mass. This has been discussed in the past, e.g., for polarons in a lattice system immersed in a continuous reservoir gas~\cite{Bruderer2007,Klein2007}. Here we are generally interested in limits where we go away from the usual regimes of polarons, but for certain limiting cases it is possible to directly extract approximate mediated interactions and effective masses for one species of atom, and comparing the visibility profile resulting from this effective model, from the full calculation including the reduction in first-order coherence due to entanglement. 

This can be approached in the limit where the species of interest (i.e., impurities) are much heavier than the second species, so that they have much lower tunnelling rate ($J_2 \ll J_1$). In this limit, we can treat the problem in a Born-Oppenheimer (B-O) approximation, which allows us to extract effective interactions as a function of distance. By considering a single impurity confined to two lattice sites, we can also determine effective tunnelling rates that reflect any increased effective mass, allowing us to generate an effective model for the behaviour of the impurities. We will do this first for a single particle of the second species per impurity, and then repeat the calculation for a case where the second species becomes the majority species.

In the B-O approximation, the full Hamiltonian~\eqref{HBH} is divided into two parts. The first part is the tunnelling of the impurities, $H_T = -\sum_{\langle l,j \rangle} J_{2}b_{2,l}^{\dag}b_{2,j}$, which is considered to be very small on the timescale relevant for the dynamics of the second species of atoms. The second part of the Hamitonian then governs the resulting configuration where the impurities are essentially taken to be motionless. With the impurities separated by a distance $R$ (in terms of lattice parameter), the corresponding Hamiltonian can be written as
\begin{align} 
H_{R}= -\sum_{\langle l,j \rangle} J_{1}b_{1,l}^{\dag}b_{1,j}+\displaystyle\sum_{l}\frac{U_{1}}{2}n_{1,l}(n_{1,l}-1) + \displaystyle\sum_{l \in i,i+R} U_{12}n_{1,l}n_{2,l} + U_2 \delta_{R0}
\,, \label{H_R} \end{align}
which is independent of the overal position on the lattice,  for periodic boundary conditions. An eigenstate of the full Hamiltonian, $|\psi\rangle$, with eigenenergy $E$, can be written in terms of the eigenfunctions of $H_R$, $|\phi_{k,R}\rangle$, i.e.
\begin{align}
|\psi\rangle = \sum_{k,R}  C_{k,R} |\phi_{k,R}\rangle.
\end{align}

Applying the full Hamiltonian on $|\psi\rangle$ and using orthonormalisation of the set of $|\phi_{k,R}\rangle$ lead to
\begin{align}
\left( \langle \phi_{k,R} | H_T | \phi_{k,R} \rangle + E_{k,R}\right) C_{k,R} + \sum_{k'\ne k,R'\ne R} \langle \phi_{k,R} | H_T | \phi_{k',R'} \rangle= E  C_{k,R}, 
\label{BO}
\end{align}
where $E_{k,R}$ are the eigenenergies of $H_R$. The B-O approximation is valid when these eigenenergies are well separated. In this case, the off-diagonal coupling terms $\langle \phi_{k,R} | H_T | \phi_{k',R'} \rangle$ can be neglected. Now, $E_{k,R}$ in the uncoupled Eq.~\eqref{BO} can be interpreted to be the long-range interaction potential between the two impurities, mediated by the majority atoms tunneling faster. We can now find the ground state of the full Hamiltonian by diagonalising Eq.~\eqref{BO} with lowest energy states of $H_R$. 

\begin{figure}[t]
\hspace{1in}
\includegraphics[width=0.75\linewidth]{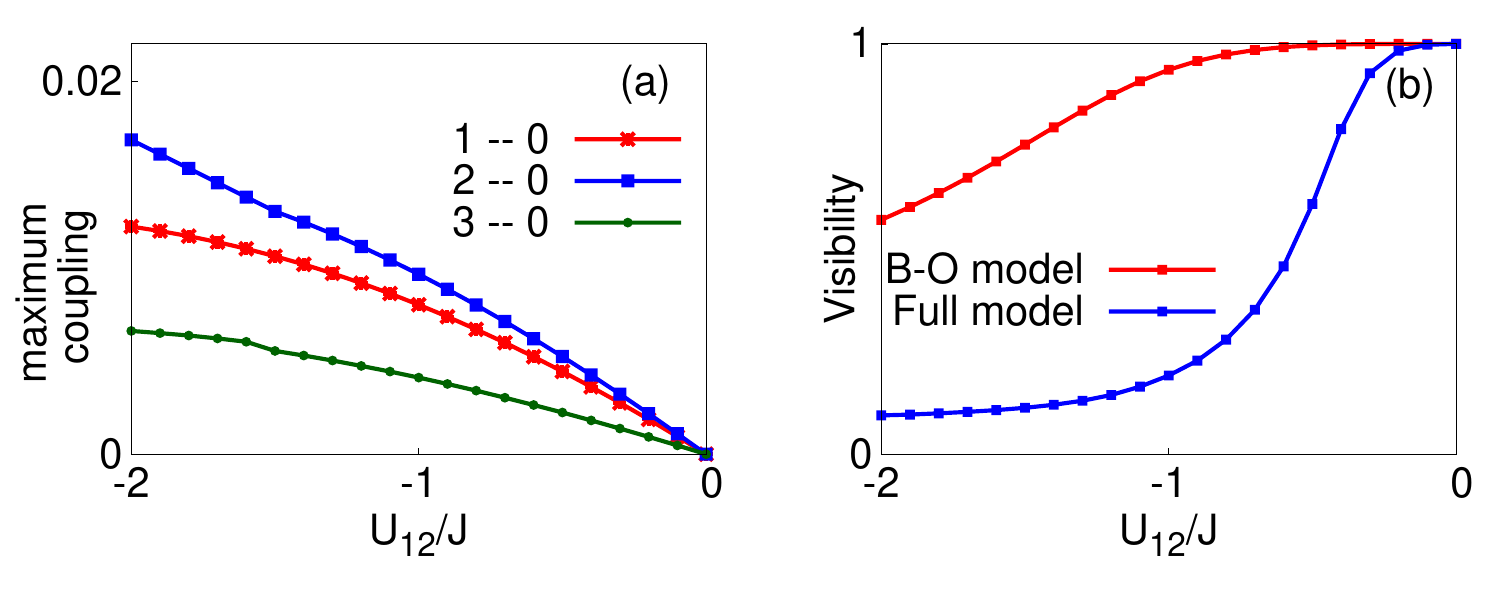}
\caption{(a) The maximum non-diagonal tunnelling coupling between the ground state (0) and $n$-th excited state, divided by their energy differences, and denoted as $n$ - - 0, for 12-site chain with 12 majority bosons, and $U_1=2J$, $J_2=0.1J$. (b) Visibility of peaks in the momentum distribution for the impurity species with interspecies interaction $U_{12}$. The red line shows the result for the B-O approximation, with $J_2=0.1J$. The blue line shows the result for the exact ground state of the same system, computed using DMRG calculations with MPS bond dimension $D=400$. \label{fig_BO_2}}
\end{figure}

We first numerically investigate the case where two atoms from each species are considered in long chains (in order to reduce finite size effects) for different values of $R$, $J_2/J_1$, $U_1$, and $U_{12}$. The findings are shown in figure~\ref{fig_BO_1} where we take a chain with $M=50$ sites and fix $J_2=0.01J$ and $U_1=2J$. The validity of the B-O approximation is checked and as shown in figure~\ref{fig_BO_1}(a), the spectrum is found to be properly gapped only for sufficiently attractive $U_{12}$. In the attractive case, as $|U_{12}|$ grows larger than $J_1$, the ground state is a bound state where the two majority particles are localized at the positions of the static impurities, with an energy gap opens to a state where only one majority particle is localized, and the other one is delocalized around that site. For repulsive $U_{12}$ the spectrum is gapless as transition to excited states is caused by any tunnelling of the lighter majority particles, therefore, we only consider attractive interactions in the following. 

The potential energy surfaces $E_R$ for this system are then computed. The attractive long-range effective interaction $U_2^{\text{eff}}$ is shown in figure~\ref{fig_BO_1}(b), as a function of the distance $R$. Here, the $U_2^{\text{eff}}$ values are the ground state energies of $H_R$, and the zero of the energy is chosen at the largest $R$. Next, we calculate the effective tunnelling $J_2^{\text{eff}}$ of an impurity in the presence of delocalised majority species bosons. This can be done by confining the impurity in a double-well in the middle of the chain with majority species bosons with small enough $U_1$. The effective tunnelling is then given by half of the energy difference between the lowest even and odd states. This is shown in figure~\ref{fig_BO_1}(c) as a function of $U_{12}$. With these parameters computed, the resulting visibility profile for two impurities is shown as the red line in figure~\ref{fig_BO_1}(d). This reflects the mediated interactions and increased effective mass captured by the B-O approximation. The blue line in figure~\ref{fig_BO_1}(d) displays the total visibility in the actual ground state of the full four particle system, which shows an additional decrease in visibility from the B-O model. 

We now carry out similar calculation for two impurities interacting with relatively larger number of majority bosons, in the context of the systems we consider in this work. In figure~\ref{fig_BO_2} we show the results for a $12$-site chain with $12$ majority species bosons and $2$ impurity bosons. For reasons stated before, we only consider attractive $U_{12}$. We also ensure that the coupling between the ground state and the excited states in the Born-Oppenheimer treatment stays small in the parameter ranges we are interested in. This is shown in figure~\ref{fig_BO_2}(a) where the maximum, taken over all $R$, of the ratio of tunnelling coupling between ground state and a few excited states and the energy difference between them, is plotted, for $J_2=0.1J$. We have taken $U_1=2J$ to ensure that the majority species is delocalised. In order to not disrupt the ground state of the majority species too much we use $|U_{12}| < U_1$. With these choices the effective parameters $J_2^{\text{eff}}$ and $U_2^{\text{eff}}$ for the two impurities are calculated as before. The resulting visibility profile is shown in figure~\ref{fig_BO_2}(b), as the blue line. To show the corresponding visibility for this system due to the combined effects, with entanglement, we compute the many-body ground state, using DMRG methods. The visibility of the impurity species is shown as the red line in figure~\ref{fig_BO_2}(b), clearly showing the additional decrease in visibility beyond what the B-O model capturing mediated interactions and an increased effective mass generates.

With this understanding we now focus on the interspecies entanglement and its effects on larger    systems beyond the two particle system. Below we will see analogous behaviour of that presented in section~\ref{Two particles on 1D lattice} in the many-body case, made somewhat more complicated by the dynamics of interacting particles in the majority species.

\section{Many-body case for impurities on a lattice}
\label{Many body problem}

In this section, we now investigate the interspecies entanglement in regimes where many-body dynamics play a key role. We identify the corresponding effects of interactions on the visibility of a $p=0$ peak in the momentum distribution, and use this to understand signatures of the many-body phase diagram of the majority species in the dynamics of the impurity atoms.

\begin{figure}[t]
\hspace{1in}
\includegraphics[width=0.75\linewidth]{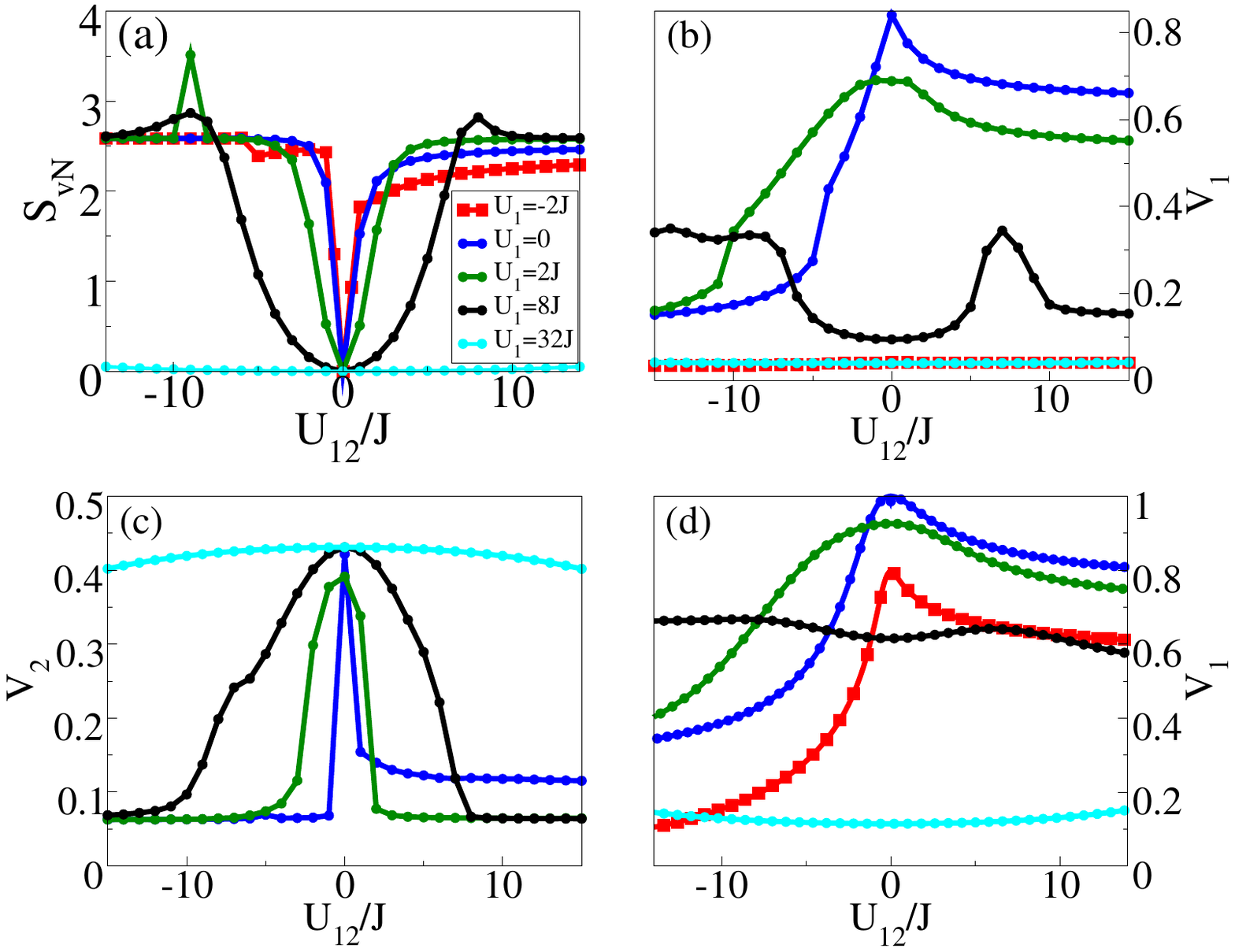}
\caption{Entanglement and momentum distribution visibility for two bosonic species on a lattice, with varying majority species interactions, as a function of interspecies interactions. (a) The von Neumann entropy of entanglement between the two species, $S_{\text{vN}}$ computed with ED methods with periodic boundary conditions for $M=6, N_1=M, N_2=M/3$ and $U_2=32J$. The sudden increase in the value of $S_{\text{vN}}$ for $U_1=2J$ is an interesting feature in the nature of the ground state arising due to interplay of the system parameters. This is discussed in detail in the main text. (b) The visibility $V_1$ of the momentum distribution peaks for the majority species $1$ as a function of interspecies interaction $U_{12}$ for a lattice chain length $M=16$ with $N_1=M, N_2=M/4$ and $U_2=32J$, computed using DMRG calculations with MPS bond dimension $D=128$ for a range of $U_1$ values. (c) The visibility of the momentum distribution peak  $V_2$ of the impurity species $2$ for the same parameters as part b. (d) The visibility from mean-field calculations in a homogeneous lattice using the bosonic Gutzwiller ansatz for the same values of $U_1$ and with $M=32$.\label{num1}}
\end{figure}

\subsection{Effects on the majority species}
For each of the cases in this section, we compute the ground states of the Hamiltonian Eq.~\eqref{HBH}, which we compute using ED (exact diagonalisation), DMRG, and Gutzwiller mean-field methods as discussed in section~\ref{model}. As discussed above, we refer to the height of the peak of the quasi-momentum distribution per particle, denoted by $V_{\sigma}$ for the species $\sigma$, as the visibility of the momentum distribution. Similarly, the von Neumann entropy $S_{\text{vN}}$ shows the effects of entanglement between the two species. For the ED calculations (with periodic boundary condition) the lattice consists of $6$ sites ($M$) and for the mean field and DMRG calculations we have used $M = 32$ and $M=16$ respectively. In all cases the number-dominant reservoir species is at unit filling ($N_1=M$) and the impurity species is at quarter filling ($N_2=M/4$), except for $M=6$ where we have taken $N_2=M/3$.

In general, as the two species become more entangled with increasing interactions, the visibility of the momentum distribution clearly decreases. However, there are a number of important many-body phenomena that are visible at specific points in the visibility profiles as a function of interspecies interaction. The passage from a delocalised phase to a localised one for species $1$, as the intra-particle interaction increases, can cause a rise in the visibility, while entanglement keeps increasing. Alternatively, there are points in the parameter space where the ground state goes through abrupt structural changes, causing a sudden increase in the von Neumann entropy, which cannot be captured by observing the momentum distribution alone. In the following we analyse these features in more detail.

In figure~\ref{num1} we show the entanglement of the species $1$ in terms of the von Neumann entropy $S_{\text{vN}}$ as a function of interspecies interaction from the ED results in figure~\ref{num1}a, and the visibility profiles from the DMRG calculations in Figs.~\ref{num1}(b), (c) and mean field calculations in figure~\ref{num1}(d). We notice an increase in entanglement and decrease in visibility as the interspecies interaction $U_{12}$ is increased from zero, as was seen in the previous section for the system of two bosons on an optical lattice. A change in $U_2$ does not have significant effect on the general entanglement or the visibility profiles, so we fix the value to be $U_2=32J$.  

When the majority species $1$ particles are non-interacting ($U_1=0$) they are delocalised at $U_{12}=0$ and we see a high visibility of the momentum peak at $p=0$ in this case. When $U_{12}$ is increased, we see the decrease in the visibility, in a form that is largely familiar from the two-particle case in the previous section. 

For repulsive interactions between majority atoms, we first look at $U_{1}=2J$ where the particles of species $1$ are still largely delocalised at $U_{12}=0$ as they still are in the superfluid regime of the single-species Bose-Hubbard model. The finite $U_1$ value, however, results in a slight decrease of the visibility $V_1$. As $U_{12}$ is increased from zero we see similar behaviour for both the visibility and von Neumann entropy to that seen in the $U_1=0$ case but now taking $U_{12}>0$  further reduces the delocalisation of species $1$ atoms. This leads to a small further decrease of visibility compared with the non-interacting case, but for $U_{12}<0$ the decrease in visibility is correspondingly less than the $U_1=0$ case. 

We see strong features of the many-body physics of the majority species entering the dynamics as we further increase $U_1$, so that $U_1/J$ is larger than the critical value for the Superfluid-Mott Insulator transition. In 1D this critical value has been reported as, $(U/J)_c \approx 3.3$~\cite{Kuhner2000}. For $U_1=8J$ we see that the visibility has a very different shape, which is characteristic now of the behaviour when the majority species is in a Mott Insulator regime for $U_{12}=0$. We see that the visibility $V_1$ here has a minimum at zero interspecies interaction. This can be understood as being due to the fact, in the Mott Insulator the particles are exponentially localised at each lattice site. This results in a broadening of their momentum distributions and causes the dip in the height of the momentum peak.  For low interspecies interaction the particles from the species $2$ do not have sufficiently strong interactions to excite species $1$ atoms out of the Mott Insulator entirely, but they do lead to some delocalisation through virtual amplitudes to create such excitations. This can also be seen in an increase of the von Neumann entropy. If we look at the $U_1=8J$ line in figure~\ref{num1}(b) the subsequent local maxima on the both sides of the minimum occurring at $U_{12}=0$ happen due the increase in $U_{12}$ where the effect of the presence of a second species becomes stronger. As $U_{12}$ becomes comparable to $U_{1}$, the energy input due to the presence of a species $2$ particle disrupts the localized phase as the energy penalty for having a double occupation of species $1$ is comparable to the energy required to put two particles from the different species on a single site. Thus the species $1$ particles begin to delocalise and the visibility increases substantially. However further increasing $U_{12}$ imposes a restriction on this delocalisation process which causes a drop in the visibility again.

\begin{figure}[t]
\hspace{1in}
\includegraphics[width=0.75\linewidth]{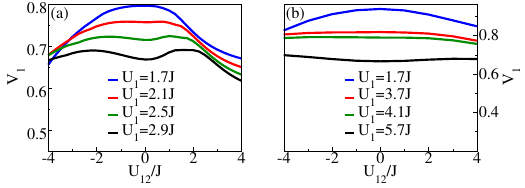}
\caption{The visibility $V_1$ of the peak in the momentum distribution for the majority species $1$, shown for different interaction strengths of the majority species as a function of interspecies interaction $U_{12}$. (a) Calculations on a lattice of length $M=16$, with $N_1=M, N_2=M/4$ and $U_2=32J$. We fix $U_2=32J$ and look at $U_1$ values somewhat less than the critical point for the Mott Insulator to superfluid transition, $(U/J)_c\approx 3.3$. We see that the behaviour of the visibility changes from going through a maximum for zero interspecies interaction to a minimum. In (b) we show the results in the same interaction parameter regimes, carried out with mean field calculations in a homogeneous lattice with $M=32$ using the bosonic Gutzwiller ansatz. As in a, we chose $N_1=M, N_2=M/4$ and $U_2=32J$. \label{num2}}
\end{figure}

The interplay between the different interaction parameters gives rise to some particularly interesting individual features at particular points in parameter space, most notably a surprisingly large increase in the value of the von Neumann entropy that occurs for $U_1=2J$ at around $U_{12}=-9J$, as can be seen in figure~\ref{num1}(a) (green line). This happens due to drastic changes in the nature of the ground states, and shows how sensitive this measure can be to such structural changes, in a regime where this cannot be detected via momentum distribution changes. Around $U_{12}=-10J$ it is energetically favourable to have all the species $1$ and species $2$ particles at one single site. In figure~\ref{num1}(a), this can happen in $6$ possible ways as we look at a $6$ site system with periodic boundary condition. On the other side of the peak-like structure, around $U_{12}=-8J$, it is energetically favourable to have both the species $2$ particles on adjacent sites, and this configuration can also achieved in $6$ different ways. The von Neumann entropy is therefore indeed $\log_2{6}$ on both sides of the peak. Now around the peak, which is near $U_{12}=-9J$ all the $12$ configurations become important and the von Neumann entropy becomes $\log_2{12}$. Carrying out a Schmidt decomposition between the species reveals that the ground state is very close to a maximally entangled states with $6$ almost equal singular values for $U_{12}=-10J$ and  $U_{12}=-8J$, and $12$ almost equal singular values for $U_{12}=-9J$. The other singular values are suppressed by at least three orders of magnitude. Looking at the energy levels of the composite system we can also see that the lowest six levels are very close to each other at $U_{12}=-10J$ and  $U_{12}=-8J$ whereas there is an avoided crossing with second lowest six levels at around  $U_{12}=-9J$. For a general $M$-site system with two impurities with large intra-species repulsion, this jump of $S_{\text{vN}}$ from $\log_2{M}$ to $\log_2{2M}$ would occur at particular value of attractive $U_{12}$, determined by the system parameters. A simple estimation of energies in the two distinct configurations shows that this happens around $|U_{12}/J| \approx U_1N_1/2 + U_2/N_1$.

For $U_1=32J$ the particles in species $1$ are in the deep Mott insulator regime and in the range of  $U_{12}$ that we are looking at here the energy input in the system by the presence of the particles of species $2$ cannot affect the Mott insulator as $U_{12}$ is always much smaller than $U_{1}$. Since varying the interaction strength does not entangle species $1$ with the other the von Neumann entropy stays at zero. The visibility of these highly site-localised species $1$ particles also stays constant at a very low value which is even much smaller for the mean field treatment as we can see in figure~\ref{num1}(d). This is because in mean field treatment the spatial correlations in a Mott insulator are exactly zero and in a numerically exact treatment they fall exponentially with the distance in space. 

In figure~\ref{num2} we look more closely at the visibility profile which changes from going through a maximum at zero interspecies interaction (for example, the $U_1=1.7J$ line in figure~\ref{num2}(a)) to a minimum (for example, the $U_1=2.9J$ line in figure~\ref{num2}(a)) as a function of the $U_1$ value. Here we notice the transition like feature which occurs in the regime where particles become more localised, but note that it occurs before the superfluid to Mott insulator phase transition in 1D, as it occurs at around $U_1=2J$ when computed using DMRG.

\subsection{Effects on the impurity species}

\begin{figure}[t]
\hspace{1in}
\includegraphics[width=0.75\linewidth]{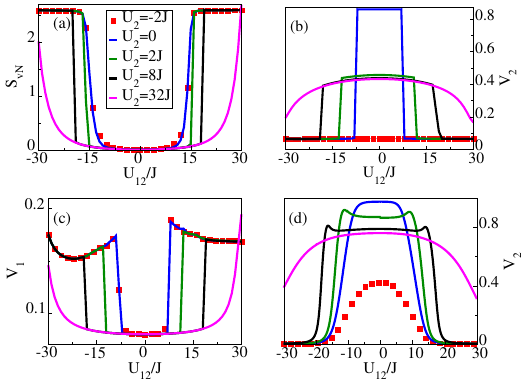}
\caption{Entanglement and momentum distribution visibility for two bosonic species on a lattice, with varying impurity species interactions, as a function of interspecies interactions. (a) The von Neumann entropy of entanglement between the two species, $S_{\text{vN}}$ from ED calculations (with periodic boundary conditions) for $M=6, N_1=M, N_2=M/3$ and $U_1=32J$. (b) The visibility $V_2$ of the peak in the momentum distribution for the impurity species $2$ as a function of interspecies interaction $U_{12}$ for a lattice chain length $M=16$, with $N_1=M, N_2=M/4$ and $U_1=32J$. This is computed from calculations with bond dimension $D=128$ for a range of $U_2$ values depicted in the legend. (c) The visibility $V_1$ of the momentum distribution peak for species $1$, with the same parameters as in b. (d) The results in the same interaction parameter regimes carried out using mean field calculations in a homogeneous lattice with the bosonic Gutzwiller ansatz for the same values of $U_2$, but with $M=32$.   \label{num3}}
\end{figure}

In figure~\ref{num3} we further investigate the effects on the impurity particles (species 2). As noted above, the properties of these impurity particles depend strongly on the many-body state of the species $1$ particles and therefore should be affected by the choice of $U_1$ values, allowing them to be used as probes for the physics of the species 1 particles. The visibility of species $2$ as a function of $U_{12}$ in general has a peak around $U_{12}=0$ that decreases on each side. This peak-like structure starts broadening as we keep increasing $U_1$ starting from $U_1=0$. The visibility profiles are quite similar to those seen in the previous section for the system of two bosons on an optical lattice in terms of the mechanisms that create a maximum at $U_{12}=0$ and a slower decrease for repulsive $U_{12}$. The value of the maximum visibility also follows a similar trend as a function of $U_2$ and falls sharply for attractive $U_2$ whereas it falls much more slowly on the repulsive side. This particular behaviour is seen in figure~\ref{num1}(c). For very large and positive $U_1$ the majority species particles are localised and for small interspecies interaction the entanglement remains very small. This phase however is disrupted at sufficiently strong $U_{12}$, causing steep rise in entanglement and subsequent decline in impurity visibility profiles.  These interesting features are reported in the following paragraphs and figure~\ref{num3}(a), (b), and (d). In figure~\ref{num3}(c) we show the visibility $V_1$ of the species $1$ for identical parameters, computed with DMRG calculations.

For $U_1=32J$ and unit filling, species $1$ particles are in the deep Mott insulator regime. For $U_{2}=0$ we expect the impurities to behave similarly to free particles on a modified lattice and as a result there is little entanglement with species $1$, causing a strong visibility around $U_{12}=0$. As  $U_{12}$ is increased, it eventually becomes energetically favourable for the impurity particles to be found together at one single lattice site and push out the species $1$ particle, creating a hole in the Mott insulator. This causes an abrupt structural change  in the localised phase of species $1$, in the sense that the impurity atoms now participate in a delocalised form in the Mott insulating phase, which will reflect the z-antiferromagnet phase of a  general two-species Bose-Hubbard model~\cite{Altman2003,Powell2009} (which will generally exhibit phase separation). The impurity particles become localised in position space by the species $1$ particles through this process and therefore we see a sharp rise in entanglement that results in a sudden decrease in the visibility of species $2$. The value of $U_{12}$ at which this happens depends on the number of impurity particles and expectedly we notice this value to be $U_1/N_2$ in figure~\ref{num3} as that would be the interspecies energy required to have one species $1$ particle and all the species $2$ particles at the same site that would match the excitation energy of the Mott insulator. Now on the attractive side of $U_{12}$ around the same magnitude ($U_1/N_2$) it also becomes energetically favourable to create a hole in the Mott insulator and to have all the impurity particles on that neighbouring site of the hole where the species $1$ particle has tunnelled to. Due to this similar localisation the two species become highly entangled and the visibility $V_2$ again falls drastically. For an infinite lattice system (which is the case when one treats the problem in mean field theory) this localisation process causes the visibility to completely vanish, as shown in figure~\ref{num3}(d). For a finite system (figure~\ref{num3}(b)) the visibility falls down and  takes a constant value that decreases as we increase the system size.

Now for  $U_{2}=2J$ the visibility at  $U_{12}=0$ will be smaller than in the previous case as the impurity particles repulsively interact among themselves. This decrease in height of the visibility persists as we increase the $U_{2}$ value but not arbitrarily as we have discussed before. For small values of  $U_{12}$  the visibility again remains unchanged and we also see the same localisation effect causing a drop in visibility at sufficiently high  $U_{12}$  as before. The magnitude of $U_{12}$ at which the drop happens increases with increase in $U_2$ as the repulsion between the impurity particles also needs to be overcome. We see this for $U_2=2J, 8J$, and $32J$ in figure~\ref{num3}(b).

For attractive $U_{2}$ we expect a similar plateau-like visibility profile but with much smaller values to start with (at and around $U_{12}=0$) and we see that in figure~\ref{num3}(b) for $U_2=-2J$, although in this case the plateau-like structure is hardly visible due to such small value of the peak. The value of visibility (for all the profiles) after the drop tends to go to $1/M$ which is the lower limit for the peak of a single particle momentum distribution on a lattice with $M$ sites (one can think about two particles on a lattice with very strong inter-particle interactions where the single particle momentum distribution is completely flat in the ground state). 

\section{Effect of the reservoir size}
\label{BEC}

Up to now we have looked at strong effects of entanglement between the two species. For certain applications we would also like to identify regimes where this entanglement can be made small. This includes proposed experiments in which the impurities can be used to measure correlation functions of the majority species~\cite{Cosco2018,Mcendoo2013}. We note that it is important that we do not reduce the coupling strength to values so small that impurities cannot be used as probes on reasonable experimental timescales. Instead, if we keep the interaction strength constant, but increase the size of the reservoir, the overall effect of the impurity on the reservoir becomes small, and this is reflected in the interspecies entanglement and corresponding observables associated with the reservoir. For example, when an impurity in an optical lattice is immersed in a Bose-Einstein condensate much larger in size compared to that of the lattice, then in the limit of weak coupling there is no visible effect of the interspecies entanglement. We note that this this limit is also a necessary (but not sufficient) requirement for making a Markov approximation when the impurity atom is coupled to a reservoir in an open quantum system~\cite{Gardiner2010, Lena2020}. It would be an interesting future study to determine the degree of entanglement between impurities and the majority species as a function of time, and how this changes with the coupling strength between markovian and non-markovian regimes, e.g., in the setting of Ref.~\cite{Cosco2018}, in which impurity atoms are used to probe a majority species of bosons in an optical lattice. 

To estimate the effect of the size of the reservoir with non-interacting species, we can consider the two-particle problem where the impurity species $2$ is confined to a small part of the lattice, with $M_2$ lattice sites, as opposed to the full length, $M_1$, and we also allow for different tunnelling rates for the two species. 

\begin{figure}[t]
\hspace{1in}
\includegraphics[width=0.75\linewidth]{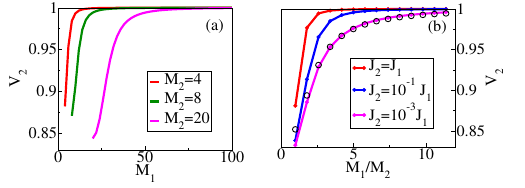} 
\caption{Effects of changing the relative lattice size or tunnelling amplitude for the impurity species and the majority species. (a) The visibility of the momentum distribution peak for a single impurity particle $V_2$ for $U_{12}=8J$, normalised to the value for $U_{12}=0$, and plotted for different $M_2$ values while increasing $M_1$. We see  that for $U_{12}=8J$, $V_2$ increases up to the non-interacting value when $M_1/M_2$ is around $3$. (b) The visibility of the impurity particle $V_2$ for $U_{12}=8J_1$ shown for different $M_1/M_2$ values while also changing the relative tunnelling $J_2/J_1$. We see that for $J_2/J_1=10^{-3}$ the calculation matches well with analytical calculations using the tunnelling term for the impurity particle as a perturbation (black circles). We have taken $M_2=5$ for this calculation. In both (a) and (b) we have used periodic boundary conditions on both lattice systems for exact diagonalisation. 
\label{5}}
\end{figure}

In figure~\ref{5}, we examine the solution to this problem, and show the characteristic differences in visibilities as a function of the ratio of the lattice sizes, and for different tunnelling rates. As can be seen in figure~\ref{5}(a), $V_2$ increases steadily for smaller values of $M_1$ for a given $M_2$ and after $M_1/M_2=3$ reaches the value obtained in the absence of interactions.

In figure~\ref{5}(b), we explore the effect of having a heavier impurity particle in the lattice system so that its tunnelling rate $J_2$ is smaller than $J_1$. In particular, we would like to find out how small $J_2$ should be compared to $J_1$, so that we can treat the tunnelling term for the impurity particles in perturbation theory. We can solve the two-particle problem both perturbatively, in the limit of small $J_2/J_1$, and also numerically. Figure~\ref{5}(b) shows the corresponding visibilities, $V_2$, as solid lines found with exact diagonalisation, against $M_1/M_2$ for descending values of $J_2/J_1$. For $J_2/J_1=10^{-3}$ we observe good agreement with results from perturbation theory.

We see that, consistent with our expectations, increasing the overall size of the majority species lattice while restricting to $U_{12}>0$ rapidly achieves a regime where the entanglement and the effects on the visibility both become small. This effect is interestingly enhanced for equal interactions, and heavy impurities require a larger lattice for the majority species to reach the same regime. 

\section{Summary and outlook}
\label{Conclusion}
In this article we have investigated the many-body entanglement between two bosonic species in an optical lattice. We quantify how the interspecies entanglement introduces additional effects, beyond mediated interactions or a changed effective mass, also in limits where it is possible to distinguish those quantities.  In particular, the change of the momentum distribution of each, as a function of interaction strength, could be used as a direct probe of the majority species, either by observation of the impurity atoms or by observation of their effects on the majority species. We have also identified regimes where the entanglement is small, which would be useful in more complex probe experiments~\cite{Mcendoo2013,Cosco2018}. 

The interspecies entanglement could also be directly measured using interference techniques with multiple copies of the state in a quantum gas microscope, alternately performing the scheme of Refs.~\cite{Islam2015,Kaufman2016,Daley2012,Moura-Alves2004} for one or both atomic species. This also opens the path towards future studies of impurity atoms and non-Markovian dynamics when they are immersed in a strongly interacting reservoir gas~\cite{Cosco2018,Lena2020}.

\ack{We thank Anton Buyskikh, Stephan Langer, and Alex Tacla for stimulating discussions. This work was supported in part by AFOSR MURI FA9550-14-1-0035, and by the European Union Horizon 2020 collaborative project QuProCS (grant agreement 641277). S.~S.~acknowledges the support of the (Polish) National Science Center Grant No.~2016/22/E/ST2/00555. D.~S.~acknowledges support from NSF PHY-1607633.}

\appendix

\section{Solution for two particles on a lattice}
\label{appendix_two}

Here we provide further details of the derivation of results for two particles on a lattice, from section~\ref{Two particles on 1D lattice}. The coefficients of the state can be written (by separating into centre of mass and relative coordinates) as 
\begin{align}
\psi(x,y)=\sqrt{\frac{1}{M}}e^{iKR}\psi_K(r)
\,,\end{align}
where the total number of sites, $M=2L+1$. With redefined effective tunnelling rate for the relative coordinate $J_K=2J\cos{(K/2)}$ and $K$-mode energy $E_K$ we now have
\begin{align}
-J_K\left(\psi_K(r+1)+\psi_K(r-1) - 2\psi_K(r) \right)+U\delta_{r,0}\psi_K(r) =E_K \psi_K(r)
\,. \label{r0} \end{align} 

In the attractive case ($U<0$) the condition for a bound solution is $(E_K-4J)<-2J_K$. Using normalization conditions and the inductive nature of Eq.~\eqref{r0} we can obtain the relative wavefunction for the lowest energy bound state. If we think of Eq.~\eqref{r0} as an equation describing a single particle on a lattice with indices running from $-L$ to $L$ then we obtain the following normalized wavefunction (for $p=0$ since we are looking for the lowest energy state),
\begin{align}
\psi(r)=\sqrt{\frac{1-e^{-2q}}{1+e^{-2q}-2e^{-q(M+1)}}} e^{-q|r|}
\,, \label{psi1} \end{align}
where  $q$ is real and is the solution of $\cos{(iq)}=\left(4J-E\right)/4J$ and the bound state energy $E=-\sqrt{U_{12}^2+16J^2}+4J$.

In the repulsive case ($U_{12}>0$) the lowest energy state can be computed making use of the periodic boundary condition along with the assumption that the wavefunction reaches maximum at the boundaries. The normalized relative wavefunction is 

\begin{align}
\psi(r) = \left\{ \begin{array}{lr} 
Ae^{ikr}+Ae^{-i2kL}e^{-ikr}  & : r \le 0 \\ 
Ae^{-i2kL}e^{ikr}+Ae^{-ikr}  & : r > 0 
\end{array} \right. 
\,, \label{psi2} \end{align}
where $A= (2(M+\cos{(2kL)}+2\text{Re}((e^{-i(M+1)k}-e^{-i2Mk} )/(1-e^{-i2k}))^{-1/2} $ and $k$ is given by $\tan{(kL+\pi/2)}+ (4J/U_{12})\sin{k}=0$. The ground state energy $E=4J(1-\cos{k})$.

Now we can also look at the limiting values of the single particle von Neumann entropy $S_{\text{vN}}=-\text{Tr}(\rho_1\text{log}_2\rho_1)$ where $\rho_1$ is the single particle reduced density matrix. For the attractive case, as $U_{12} \to -\infty$, we have $e^q \to 0$ and $\psi(x,y) \to \sqrt{1/M} \delta_{x,y}$. Therefore we can show $S_{\text{vN}} \to \log_2{M}$. For the repulsive case as $U_{12}/J \to \infty$, we have $k \to \pi/2L$ and  $\psi(x,y) \to \sqrt{1/M} \sin{\left(\pi |x-y| / 2L\right)}$. In this case we have
\begin{align}
S_{\text{vN  }} \to - \displaystyle\sum_{x,x'} \left( \frac{\displaystyle\sum_y \sin{\left(k |x-y|\right)} \sin{\left(k |x'-y|\right)} }{M}  \right) \log_2{\left( \frac{\displaystyle\sum_y \sin{\left(k |x-y|\right)} \sin{\left(k |x'-y|\right)}}{M}  \right)}
\,. \end{align}
This is the same as the limiting result shown in figure~\ref{two}.

\section*{References}
\bibliographystyle{iopart-num} 
\bibliography{interspecies_entanglement_impurity}

\end{document}